\documentclass{article}

\usepackage{graphicx}

\begin{document}

\title{Applied-Field Effects on Benzene Transmission}

\date{August 8, 2016}

 \author{Sydney G. Davison\thanks{email: sgdaviso@uwaterloo.ca}
 \thanks{Department of Applied Mathematics, University of Waterloo, Waterloo, ON, N2L 3G1, Canada} \thanks{Department of Physics and the Guelph-Waterloo Physics Institute, University of Waterloo Campus, Waterloo, ON, N2L 3G1, Canada}
 \and Kenneth W. Sulston\thanks{email: sulston@upei.ca; corresponding author} 
\thanks{School of Mathematical and Computational Sciences, University of Prince Edward Island, Charlottetown, PE, C1A 4P3, Canada} }

\maketitle

\section{Abstract}

By expressing the discrete Schr\"odinger equation as a second-order finite-difference equation with constant coefficients,
the renormalization equations for substituted benzene dimers are derived via the $c_n$-coefficient elimination procedure.
On subjecting the benzene molecule to a linear applied field, the resulting field-modified site energies are obtained, by
projecting the site-energy locations onto a corresponding benzene dimer axis.
Incorporating these modified site energies into the Lippmann-Schwinger scattering theory, enables the field effect
to manifest itself in the substituted benzene electron transmission spectral function $T(E)$.
Variations in the $T(E)$ energy spectra, arising from increases in the applied field gradient $f$, are described for
each substituted benzene, and comparison made between their various patterns' behaviours.
A common feature of the $T(E)$ curves is their shifts to lower energies, as $f$ increases to a 
calculated limiting value.

\section{Introduction}

Interest in the effects of an applied electric field dates back to the pioneering work of Zener \cite{ref1} in the 1930's,
when he investigated the electrical breakdown in solid dielectrics.
Subsequent theoretical work \cite{ref2}-\cite{ref4} established the fact, rather later, that the presence of a linear
electric field resulted in the existence of a tilted-band of discrete electron energy levels, in the  case of a 
linear atomic chain.
A brief review of this period has appeared in the Green-function treatment of the subject. \cite{ref5a, ref5}

By contrast, molecular electronics emerged more recently in 1974, when Aviram and Ratner \cite{ref6}
first proposed that molecules could play the role of active components in devices.
In the realm of miniaturization, single molecules have  a distinct advantage over quantum dots \cite{ref7}
single-electron transistors, because the electronics of molecular devices can be chemically ``designed'' to
suit specific applications.

In molecular electronics, a three-terminal device is the preferred choice for many applications.
However, for it to be a possible alternative to the metal-oxide-semiconductor field-effect transistor,
the gate voltage, at a fixed small drain-source bias, must be able to amplify the current by orders of magnitude.
In addressing this situation, Di Ventra {\it et al} \cite{ref8} reported their work on a parameter-free, fully
quantum mechanical, transport calculation of a three-terminal molecular device, namely, a benzene-1,4-dithiolate
molecule attached to two electrodes and a capacitive gate.
Their results showed that the molecule's resistance rose from its zero-gate-bias value to a value approximately
equal to the quantum of resistance of 12.9 k$\Omega$, when resonance tunnelling via the $\pi^*$ anti-bonding
orbital occurs.

Using benzene as a prototypical example, Hettler {\it et al} \cite{ref9} investigated the novel effects arising
when the transport through several competing electron configurations becomes possible.
The transport calculations were performed in the weakly-coupled regime and an effective model extracted
from the molecule's electronic structure calculations.
It was assumed that the benzene transport was dominated by the $\pi$-electron system, whose molecular
states were subjected to an applied electric field.
When the electrodes were coupled to the para-benzene positions, a current collapse and strong negative
differential conductance were predicted, due to a ``blocking'' state, while in the meta-benzene situation
the I-V curve was found to have a series of steps.

An interesting and important development in transmission studies of benzene was the realization that  the
molecule's molecular orbitals (MOs) and their energy spectra were modified by the presence of an electric
field \cite{ref10,ref11}.
Since the MOs contain all the quantum mechanical information in the benzene's electronic structure, and 
also provide a spatial region for the traversing electrons, it is clear that external-field modification of the
MOs opens a way of gaining useful insight into the molecule's transport properties, where the current 
flowing through is given by the Landauer-B\"uttiker formula \cite{ref12}, which utilizes the transmission
probability function $T(E)$, which is encountered later in this article.

\section{Renormalization via Coefficient Elimination}

\begin{figure}[hp]
\includegraphics[width=12cm]{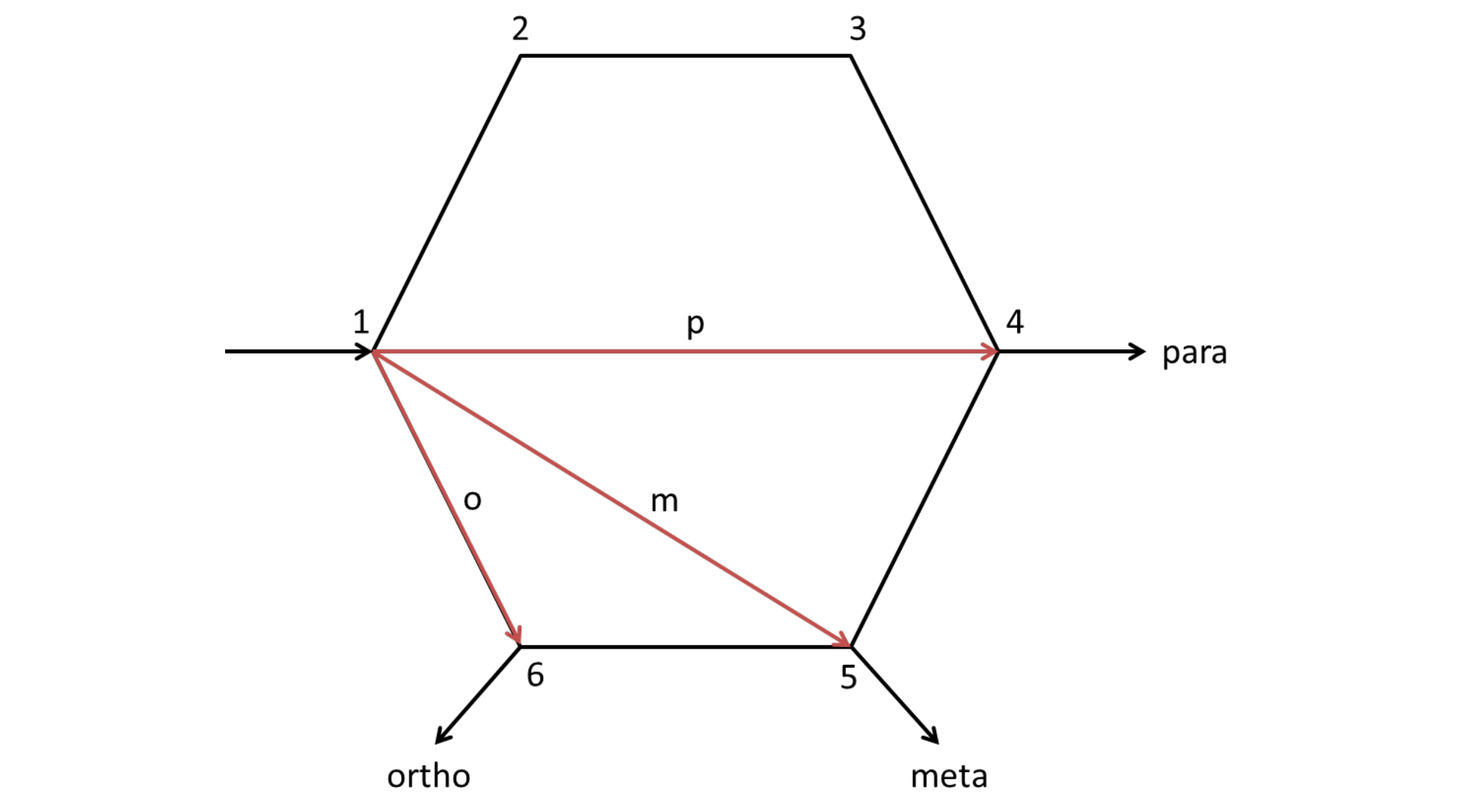}
\caption{Location of atomic-wire leads at the benzene atomic sites p(1,4), m(1,5) and o(1,6) identifying the corresponding benzene configurations.}
\label{fig1}
\end{figure}

In an earlier article \cite{ref13}, we derived the required renormalization equations via a Greenian-matrix
version of the discrete Schr\"odinger equation, which gave rise to a set of general equations describing
the decimation-renormalization procedure.
Here, we adopt an alternative route based on the well-known H\"{u}ckel molecular-orbital method \cite{ref14},
where the one-electron discrete Schr\"odinger equation is written as a second-order finite difference equation
with constant coefficients, in which the $c_n$-coefficient elimination scheme is akin to the aforementioned
decimation process.
Such an approach leads directly to a particular set of normalization equations for each of the so-called
substituted benzenes referred to as the para(p)-, meta(m)- and ortho(o)-benzene dimers (Figure \ref{fig1}).

\begin{figure}[hp]
\includegraphics[width=12cm]{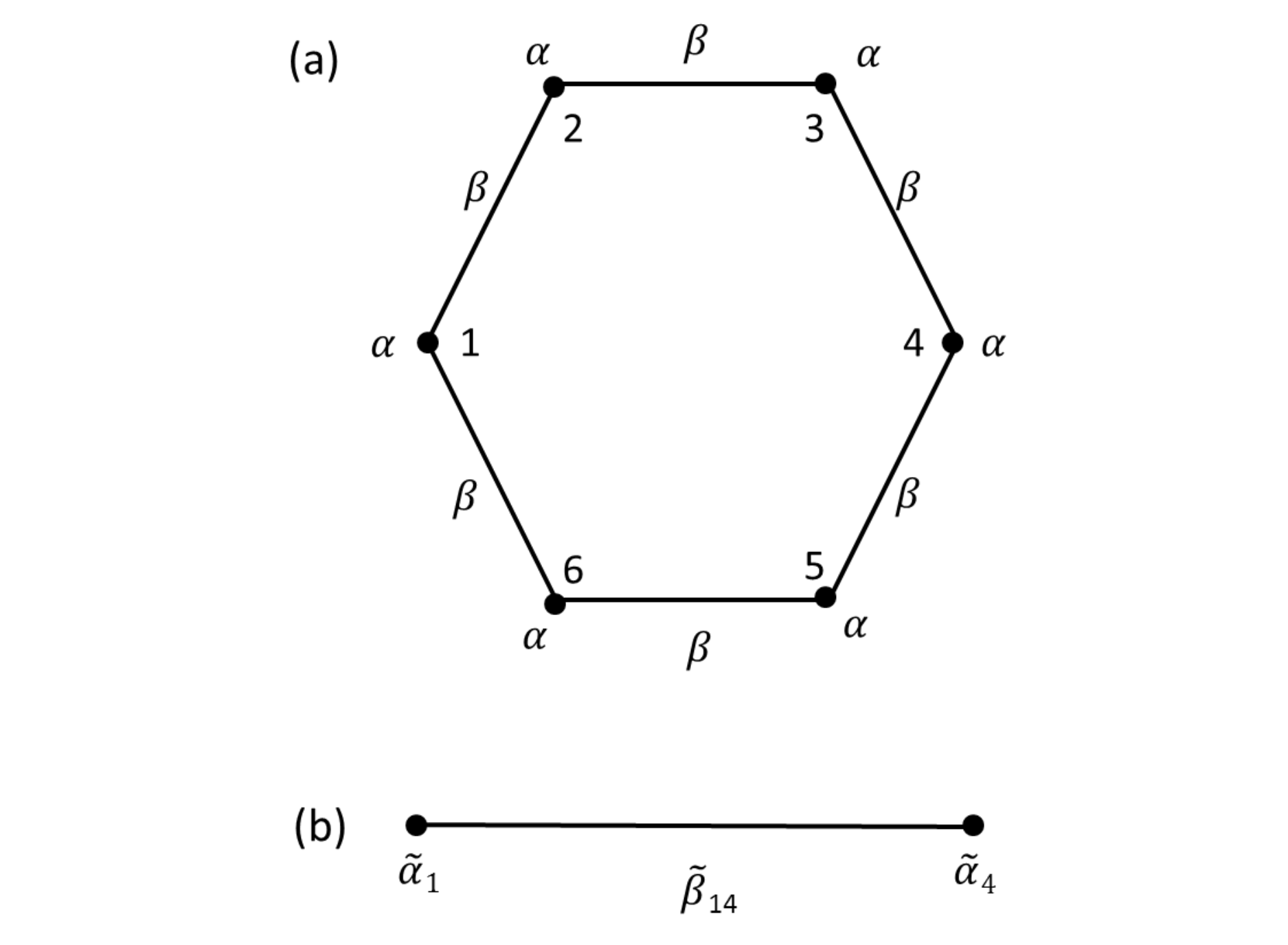}
\caption{(a) The hexagonal benzene molecule showing the six identical atomic-site energies $\alpha$ and bond energies $\beta$. (b) The  p-benzene dimer of renormalized atomic-site energies $\tilde{\alpha}_1$ and $\tilde{\alpha}_4$, with renormalized bond energy 
$\tilde{\beta}_{14}$.}
\label{fig2}
\end{figure}

To begin the $c_n$-coefficient elimination process, we cast the Schr\"odinger equation in the form of a general
nearest-neighbour difference equation for the benzene molecule displayed in Figure \ref{fig2}(a), viz.\cite{ref14},
\begin{equation}
(E - \alpha) c_n = \beta (c_{n+1} + c_{n-1}) ,
\label{eq1}
\end{equation}
E being the electron energy and $c_n$ the wave-function coefficient at the $n$-th atomic site, while $\alpha$ ($\beta$)
is the site (bond) energy of the benzene molecule. 
(Note that in the case under consideration in this section, namely $f=0$, all site energies $\alpha$ are equal. In the next section,
where we examine the case $f \ne 0$, we will require different site energies $\alpha_1$, \ldots, $\alpha_6$, in which situation
the renormalization process follows a similar, but more general, line.)
On writing equation (\ref{eq1}) out for each of the six atomic sites in the benzene molecule in Figure \ref{fig2}(a), we
obtain the set of equations
\begin{equation}
X c_1 =  c_2 + c_6 ,
\label{eq2}
\end{equation}
\begin{equation}
X c_2 =  c_1 + c_3 ,
\label{eq3}
\end{equation}
\begin{equation}
X c_3 =  c_2 + c_4 ,
\label{eq4}
\end{equation}
\begin{equation}
X c_4 =  c_3 + c_5 ,
\label{eq5}
\end{equation}
\begin{equation}
X c_5 =  c_4 + c_6 ,
\label{eq6}
\end{equation}
\begin{equation}
X c_6 =  c_1 + c_5 ,
\label{eq7}
\end{equation}
where the dimensionless reduced energy is 
\begin{equation}
X  =  (E - \alpha)/\beta .
\label{eq8}
\end{equation}

\noindent {\large {\bf A. p-benzene dimer (1,4)}} \cite{ref13}

\vspace{10pt}

As an illustrative example, we derive the renormalization equations for the case of the p-benzene dimer in Figure \ref{fig2}(b),
which resides between the (1,4) sites, where $\tilde{\alpha}_1$ and $\tilde{\alpha}_4$, with $\tilde{\beta}_{14}$, are the
renormalized site and bond energies, respectively. 
In the above equations, we seek to obtain expressions for $c_1$ and $c_4$ only, by  elimination of the other coefficients.
From (\ref{eq3}) and (\ref{eq4}), we have, respectively,
\begin{equation}
 c_2 =  (c_1 + c_3)/X ,
\label{eq9}
\end{equation}
\begin{equation}
 c_3 =  (c_2 + c_4)/X ,
\label{eq10}
\end{equation}
from which (\ref{eq10}) in (\ref{eq9}) gives
\begin{equation}
 c_2 = [c_1 +  (c_2 + c_4)/X ]/X ,
\label{eq11}
\end{equation}
so
\begin{equation}
 c_2 = (1-X^{-2})^{-1}  (X^{-1} c_1 + X^{-2} c_4) .
\label{eq12}
\end{equation}
Conversely, (\ref{eq9}) in (\ref{eq10})  yields
\begin{equation}
 c_3 = [ (c_1 + c_3)/X  + c_4]/X ,
\end{equation}
i.e.,
\begin{equation}
 c_3 = (1-X^{-2})^{-1}  (X^{-2} c_1 + X^{-1} c_4) .
\label{eq13}
\end{equation}
By dint of the symmetry in Figure \ref{fig2}(a), we see that 
\begin{equation}
 c_6 = (1-X^{-2})^{-1}  (X^{-1} c_1 + X^{-2} c_4) ,
\label{eq14}
\end{equation}
and
\begin{equation}
 c_5 = (1-X^{-2})^{-1}  (X^{-2} c_1 + X^{-1} c_4) .
\label{eq15}
\end{equation}
Hence, (\ref{eq12}) and (\ref{eq14}) in (\ref{eq2}) provides
\begin{equation}
X c_1 = 2 (1-X^{-2})^{-1}  (X^{-1} c_1 + X^{-2} c_4) .
\label{eq16}
\end{equation}
Inserting (\ref{eq13}) and (\ref{eq15}) in (\ref{eq5}), we find
\begin{equation}
X c_4 = 2 (1-X^{-2})^{-1}  (X^{-2} c_1 + X^{-1} c_4) .
\label{eq17}
\end{equation}
Proceeding further with (\ref{eq16}), we find
\begin{equation}
[X -2 X^{-1} (1-X^{-2})^{-1} ] c_1 = 2  X^{-2} (1-X^{-2})^{-1} c_4,
\end{equation}
whereby
\begin{equation}
 X [1 -2 (X^{2}-1)^{-1} ] c_1 = 2  (X^{2}-1)^{-1} c_4 .
\label{eq18}
\end{equation}
In terms of (\ref{eq8}), we can write (\ref{eq18}) as
\begin{equation}
[E -  \alpha - 2\beta X (X^{2}-1)^{-1} ] c_1 = 2 \beta (X^{2}-1)^{-1} c_4 ,
\label{eq19}
\end{equation}
which is the equation for the equivalent dimer in Figure \ref{fig2}(b), i.e.,
\begin{equation}
(E -  \tilde{\alpha}_1 ) c_1 =  \tilde{\beta}_{14} c_4 .
\label{eq20}
\end{equation}
In view of the symmetry of the benzene molecule in  Figure \ref{fig2}(a), comparing
equations  (\ref{eq19}) and (\ref{eq20}) reveals that
\begin{equation}
\tilde{\alpha}_p =  \tilde{\alpha}_1 =  \tilde{\alpha}_4 = \alpha + \tilde{\beta}_{14} X ,
\label{eq21}
\end{equation}
where
\begin{equation}
\tilde{\beta}_p =  \tilde{\beta}_{14} =  2\beta (X^2-1)^{-1} ,
\label{eq22}
\end{equation}
are the {\em renormalization equations} for the p-benzene dimer in accord with \cite{ref13}.

Likewise, equations (\ref{eq2}) to (\ref{eq8}) enable the corresponding renormalization equations to be
derived for the m- and o-benzene dimers in Figure \ref{fig1}. 
However, having provided a detailed derivation for the p-benzene dimer case, only the final results
are given for the m- and o-benzene dimers.

\vspace{10pt}

\noindent {\large {\bf B. m-benzene dimer (1,5)}} \cite{ref13}

\begin{equation}
\tilde{\alpha}_m =  \tilde{\alpha}_1 =  \tilde{\alpha}_5 = \alpha +\beta X^{-1} + \tilde{\beta}_{15}  ,
\label{eq23}
\end{equation}
\begin{equation}
\tilde{\beta}_m =  \tilde{\beta}_{15} =  \beta X^{-1} (X^2-1) (X^2-2)^{-1} .
\label{eq24}
\end{equation}

\vspace{10pt}

\noindent {\large {\bf C. o-benzene dimer (1,6)}} \cite{ref13}

\begin{equation}
\tilde{\alpha}_o =  \tilde{\alpha}_1 =  \tilde{\alpha}_6 = \alpha +\beta (X^2-2)(X^2-X-1)^{-1} - \tilde{\beta}_{16}  ,
\label{eq25}
\end{equation}
\begin{equation}
\tilde{\beta}_o =  \tilde{\beta}_{16} =  \beta (X^2-1) (X^2-2) [(X^2-1)^2-1]^{-1}   .
\label{eq26}
\end{equation}

\vspace{10pt}
We note that, because each pair of dimer site-energies are equal, all of the p-, m- and o-dimers are symmetrical,
when no applied field is present.
We should also point out that the site-coefficient elimination process, in the difference-equation approach,
parallels that of the decimation procedure in establishing the renormalization equations, which constitute the final common goal.

\section{Field-Modified Benzene}

\noindent {\large {\bf A. $\bf{p_f}$-benzene}}

\vspace{10pt}

\begin{figure}[htbp]
\includegraphics[width=12cm]{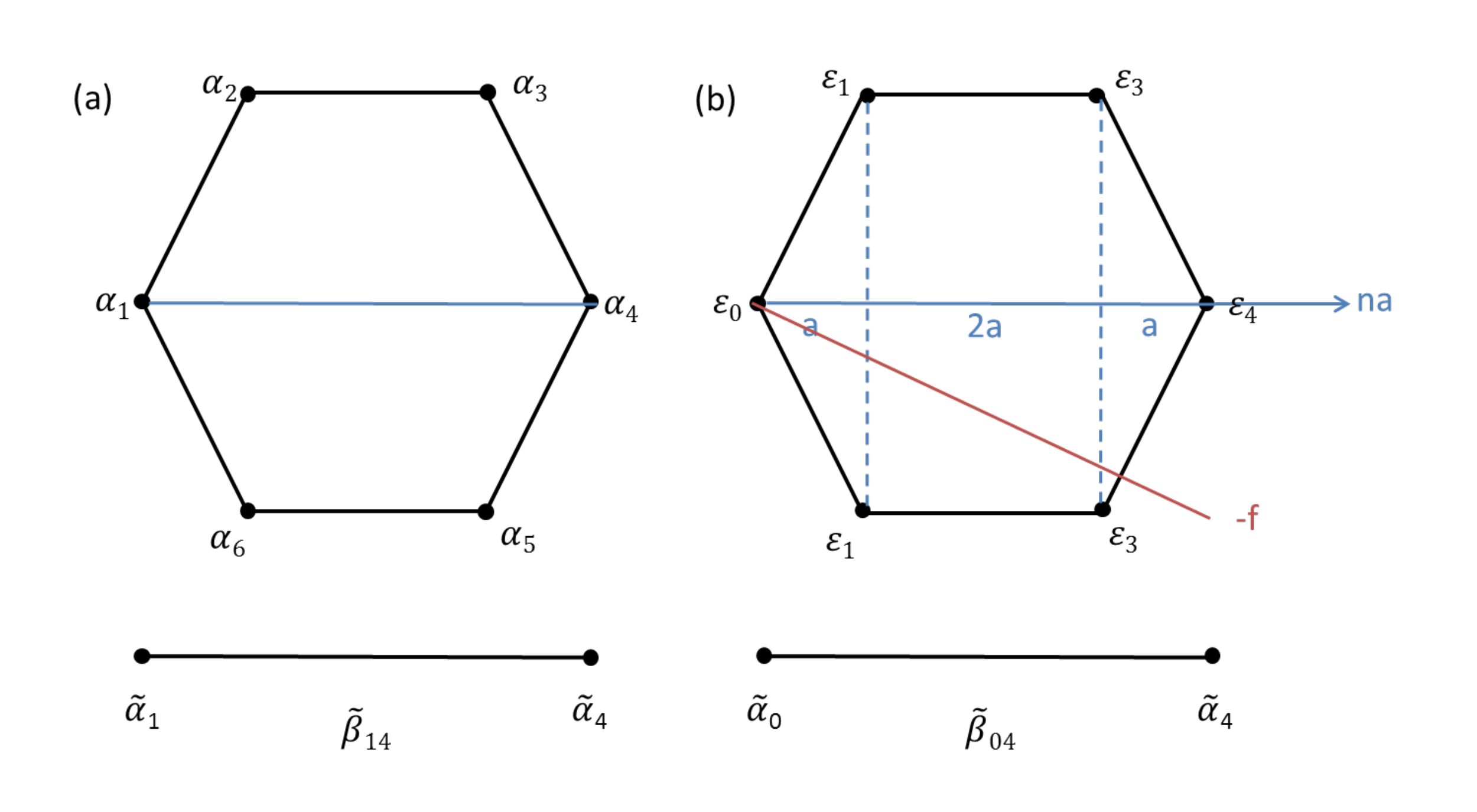}
\caption{(a) Zero-field $\alpha_n$-site locations for p-benzene, and p-benzene dimer with site (bond) energies $\tilde{\alpha}_{1,4}$ 
($\tilde{\beta}_{14}$). (b) Field-modified $\epsilon_n$-site locations for $\rm{p_f}$-benzene, as identified by their projections onto the $na$-axis, with
$\rm{p_f}$-benzene dimer site (bond) energies $\tilde{\alpha}_{0,4}$($\tilde{\beta}_{04}$).}
\label{fig3}
\end{figure}

Turning to the question of  the applied electric field, as portrayed in Figure \ref{fig3}(b), we see that the $\rm{p_f}$-benzene
dimer also provides a spatial $na$-axis between the two atomic-wire leads, at the (0,4) sites, across which a bias voltage
is established that creates a linear field of gradient strength $-f$ over the molecule, which modifies the site energies $\alpha$,
while leaving the bond energies $\beta$ unaffected.
The presence of the field manifests itself at the $n$-site energy $\epsilon_n$ via the linear relation (see, e.g., \cite{ref5})
\begin{equation}
\epsilon_n=  \alpha-n \Gamma ~,~ \Gamma=af,
\label{eq27}
\end{equation}
$\Gamma$ being the gradient of the field-induced potential, and $2a$ the spacing between the benzene atoms.
The $\epsilon_n$-site energy of the $n$-th atom in the benzene molecule is now assigned according to its location
projected onto the $na$-axis in Figure \ref{fig3}(b).
Comparing Figures \ref{fig3}(a) and (b), we see that they are linked by the relabelling scheme
\begin{equation}
\alpha_1 \to \epsilon_0~,~ \alpha_2 \to \epsilon_1~,~ \alpha_3 \to \epsilon_3~,~ \alpha_4 \to \epsilon_4~,~ \alpha_5 \to \epsilon_3~,~
\alpha_6 \to \epsilon_1.
\label{eq28}
\end{equation}
Thus, utilizing (\ref{eq28}), we can repeat the renormalization process of Section 3, although we omit the details here,
resulting in the field-modified version of equations (\ref{eq21}) and (\ref{eq22}), namely,
\begin{equation}
\tilde{\alpha}_0 =  \epsilon_0 + 2 \beta X_3 \delta_3^{-1},
\label{eq29}
\end{equation}
\begin{equation}
\tilde{\alpha}_4 =  \epsilon_4 + 2 \beta X_1 \delta_3^{-1},
\label{eq30}
\end{equation}
\begin{equation}
\tilde{\beta}_{04} =  2 \beta \delta_3^{-1} ,
\label{eq31}
\end{equation}
where
\begin{equation}
\delta_n = X_n X_1 -1 ,
\label{eq32}
\end{equation}
as the renormalization equations for the $\rm{p_f}$-benzene dimer.
In these equations also arises the $n$-site reduced energies
\begin{equation}
X_n = (E-\alpha_n)/\beta = X+nF ~,~ F=\Gamma/\beta ,
\label{eq33}
\end{equation}
$F$ being the reduced potential field gradient, and noting that $X_0 \equiv X$.
Comparing  equations (\ref{eq29}) and (\ref{eq30}) reveals that $\tilde{\alpha}_0 \ne \tilde{\alpha}_4$,
whence the $\rm{p_f}$-benzene dimer in Figure \ref{fig3}(b) is asymmetric.
Thus, the presence of the applied field has destroyed the symmetry of the zero-field p-dimer in Figure \ref{fig3}(a).
Such situations are also encountered in the following $\rm{m_f}$- and $\rm{o_f}$-benzene cases, where again only
the basic details are provided.

\vspace{10pt}

\noindent {\large {\bf B. $\bf{m_f}$-benzene}}

\vspace{10pt}

\begin{figure}[htbp]
\includegraphics[width=12cm]{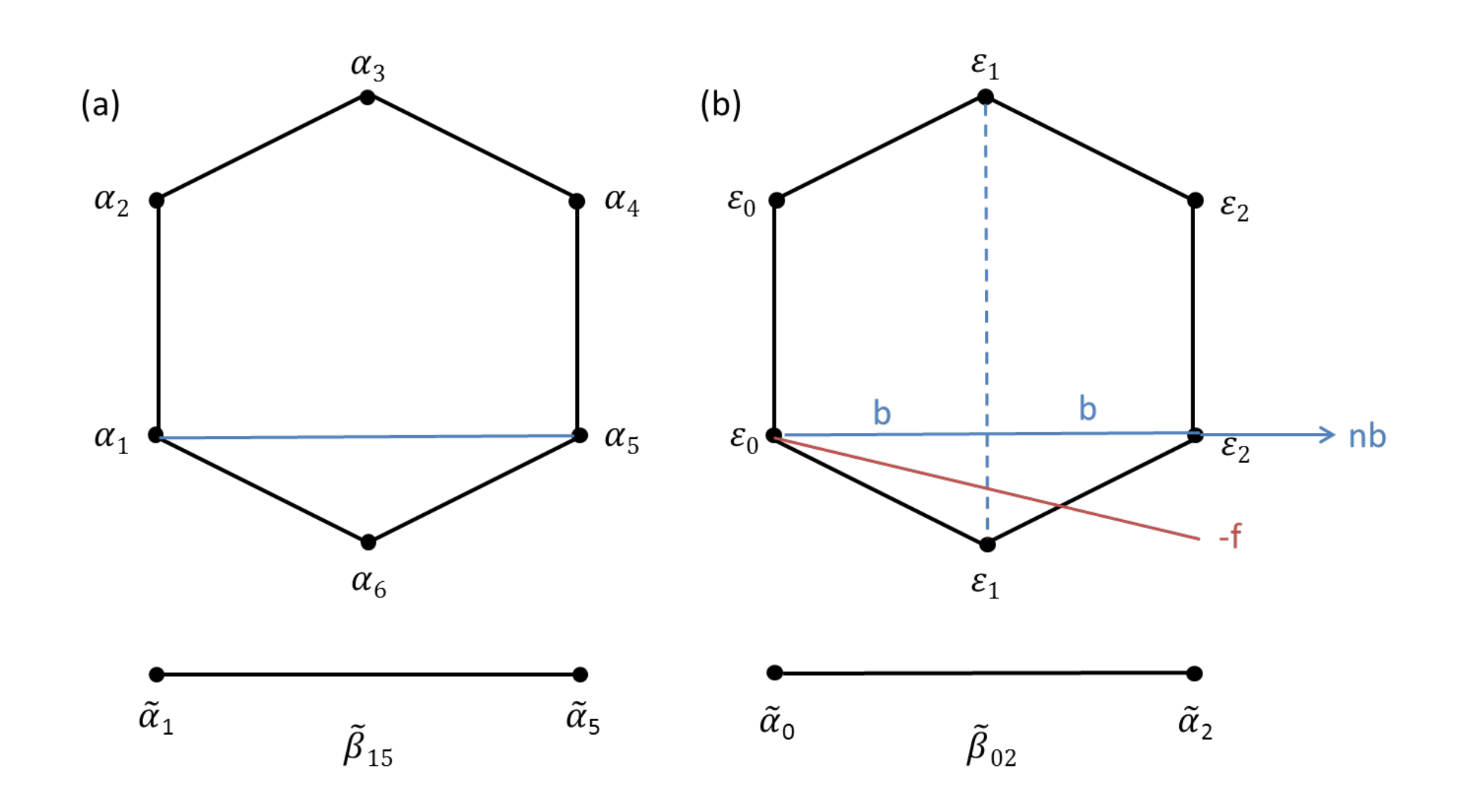}
\caption{(a) Zero-field $\alpha_n$-site locations for m-benzene. (b) Field-modified $\epsilon_n$-site locations
for $\rm{m_f}$-benzene, as identified by their projections onto the $nb$-axis. Corresponding dimer is shown in both cases.}
\label{fig4}
\end{figure}

Referring to Figure \ref{fig4}, equation (\ref{eq27}) now reads
\begin{equation}
\epsilon_n=  \alpha-n \Gamma ~,~ \Gamma=bf = \sqrt 3 af ,
\label{eq34}
\end{equation}
while the relabelling scheme becomes
\begin{equation}
\alpha_1 \to \epsilon_0~,~ \alpha_2 \to \epsilon_0~,~ \alpha_3 \to \epsilon_1~,~ \alpha_4 \to \epsilon_2~,~ \alpha_5 \to \epsilon_2~,~
\alpha_6 \to \epsilon_1.
\label{eq35}
\end{equation}
Corresponding to  (\ref{eq23}) and (\ref{eq24}), the $\rm{m_f}$-benzene dimer relations, via
(\ref{eq35}), are found to be
\begin{equation}
\tilde{\alpha}_0 =  \epsilon_0 + \beta \delta_2 (X_0 \delta_2 - X_2)^{-1} + \beta X_1^{-1},
\label{eq36}
\end{equation}
\begin{equation}
\tilde{\alpha}_2 =  \epsilon_2 + \beta \delta_0 (X_2 \delta_0 - X_0)^{-1} + \beta X_1^{-1},
\label{eq37}
\end{equation}
\begin{equation}
\tilde{\beta}_{02} =  \beta (X_2 \delta_0 - X_0)^{-1} + \beta X_1^{-1},
\label{eq38}
\end{equation}
where $\delta_n$ is given by  (\ref{eq32}).

As in the p-dimer case, the m-dimer for $f=0$ is symmetric ($\tilde{\alpha}_1 = \tilde{\alpha}_5$), while the 
$\rm{m_f}$-dimer, for $f \ne 0$,  is asymmetric  ($\tilde{\alpha}_0 \ne \tilde{\alpha}_2$).

\vspace{10pt}

\noindent {\large {\bf C. $\bf{o_f}$-benzene}}

\vspace{10pt}

\begin{figure}[htbp]
\includegraphics[width=12cm]{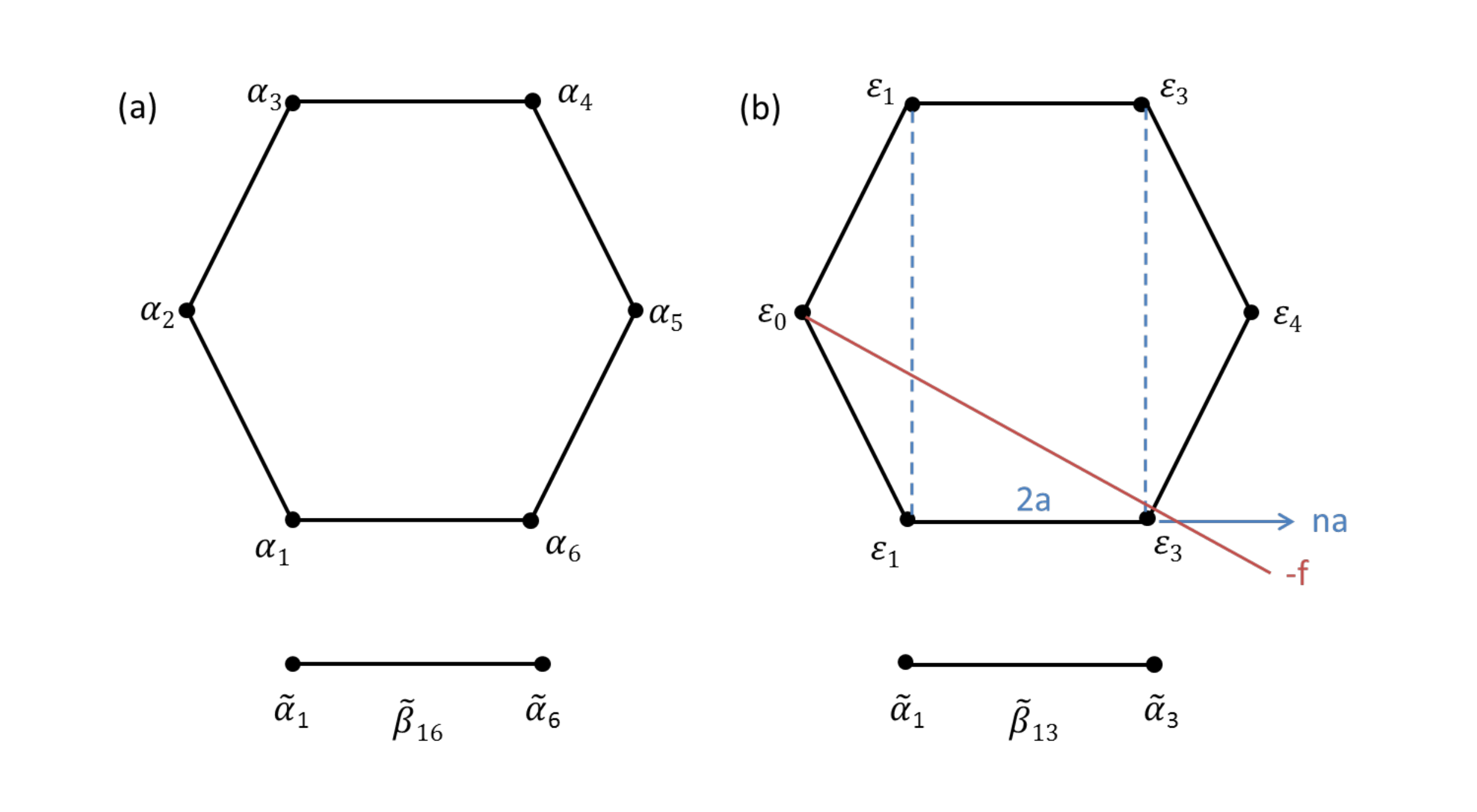}
\caption{(a) Zero-field $\alpha_n$-site locations for o-benzene. (b) Field-modified $\epsilon_n$-site locations
for $\rm{o_f}$-benzene, as identified by their projections onto the $na$-axis. Corresponding dimer is shown in both cases.}
\label{fig5}
\end{figure}

From Figure \ref{fig5}, the relabelling scheme is
\begin{equation}
\alpha_1 \to \epsilon_1~,~ \alpha_2 \to \epsilon_0~,~ \alpha_3 \to \epsilon_1~,~ \alpha_4 \to \epsilon_3~,~ \alpha_5 \to \epsilon_4~,~
\alpha_6 \to \epsilon_3.
\label{eq39}
\end{equation}
The field effect on the $n$-th site in Figure \ref{fig5}(b) is given for $\epsilon_n$ by equation (\ref{eq27}).

In the field-modified case of $f \ne 0$ in Figure \ref{fig5}(b), the equations for the $\rm{o_f}$-dimer,
corresponding to  (\ref{eq25}) and (\ref{eq26}), read
\begin{equation}
\tilde{\alpha}_1 =  \epsilon_1 + \beta X_0^{-1} [1+X_0^{-1}(\Delta_- - \Delta_+^{-1})^{-1}] ,
\label{eq40}
\end{equation}
\begin{equation}
\tilde{\alpha}_3 =  \epsilon_3 + \beta  X_4^{-1} [1+X_4^{-1}(\Delta_+ - \Delta_-^{-1})^{-1}],
\label{eq41}
\end{equation}
\begin{equation}
\tilde{\beta}_{13} =  \beta [X_0^{-1} X_4^{-1} \Delta_-^{-1} \Delta_+^{-1} (1-\Delta_-^{-1} \Delta_+^{-1})^{-1}+1]   ,
\label{eq42}
\end{equation}
where 
\begin{equation}
 \Delta_+  =X_3 - X_4^{-1} ~,~   \Delta_- = X_1 - X_0^{-1}  ,
\label{eq43}
\end{equation}
with $X_n$ given in  (\ref{eq33}).
We observe that $\tilde{\alpha}_1 \ne \tilde{\alpha}_3$, which shows that the $\rm{o_f}$-dimer is asymmetric
in Figure \ref{fig5}(b).
Hence, as with the $\rm{p_f}$- and $\rm{m_f}$-dimers, the applied field has broken the symmetry of the $\rm{o_f}$-dimer.

\section{Benzene-Leads Energy Spectra}

In this section, we look at the position of the electronic energy spectrum of the benzene molecule relative to
that of the leads, with the goal of determining how strong the applied field can be.

\vspace{10pt}
\noindent {\large {\bf A. Zero field}}
\vspace{10pt}

\begin{figure}[htbp]
\includegraphics[width=12cm]{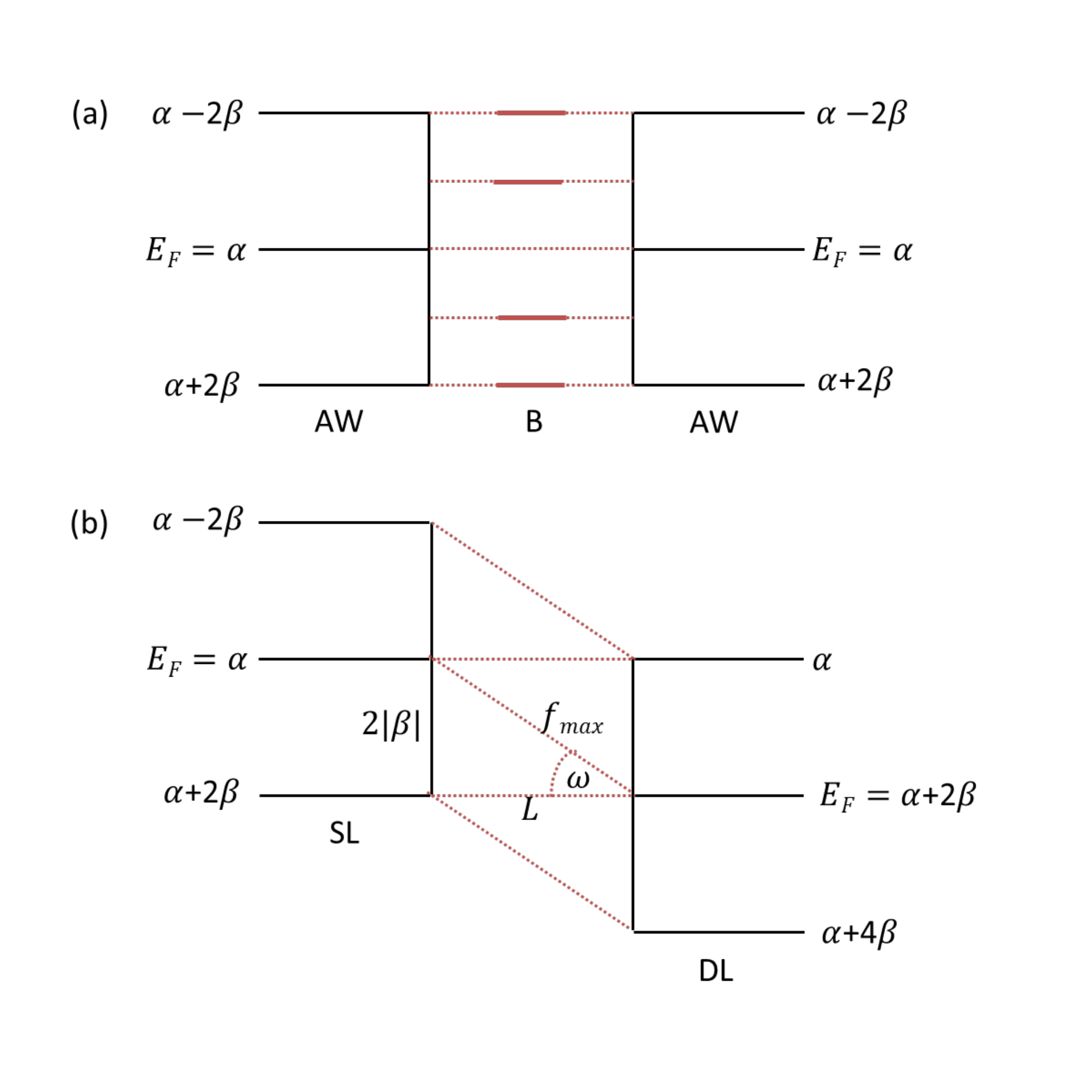}
\caption{(a) For $f=0$, aligned energy spectra of atomic wires (AW) and benzene (B).
 (b) For $f \ne 0$, tilted-band energy levels between source lead (SL) and drain lead (DL).  }
\label{fig6}
\end{figure}

In the electron-transmission studies of a benzene molecule under investigation here, atomic-wire leads
are attached to two atomic sites of the substituted benzene. 
As we shall see, in the zero-field ($f=0$) case, the benzene's and the leads' energy spectra are aligned,
as shown in Figure \ref{fig6}(a).

Each of the atomic-wire leads are represented by a linear chain of atoms, whose electronic energy
levels are derived via the tight-binding approximation.
In this nearest-neighbour calculation, the continuous energy spectrum is given by \cite{ref15} 
\begin{equation}
 E_k = \alpha +2\beta \cos \theta_k ,
\label{eq44}
\end{equation}
according to which the leads have band edges at $\alpha \pm 2 \beta$ and Fermi levels at $E_F=\alpha$.

Meanwhile, the well-known H\"{u}ckel treatment of benzene \cite{ref14} yields energy levels at 
$\alpha \pm  \beta$ (doubly degenerate) and $\alpha \pm 2 \beta$, which gives rise to the discrete
energy-level line spectrum in  Figure \ref{fig6}(a), where the upper and lower levels are found at the
same locations as the leads' band edges. 
Attachment of the leads to the benzene has the effect of broadening and shifting
these discrete levels, and possibly breaking the degeneracy (not shown shown in Figure \ref{fig6}(a)).
Nonetheless, the basic picture of Figure \ref{fig6}(a) holds true, indicating an alignment of the benzene
energy levels with those of the leads.

\vspace{10pt}
\noindent {\large {\bf B. Non-zero field}}
\vspace{10pt}

By establishing a bias voltage between the leads, a linear electric field of 
gradient strength $-f ~(\ne 0)$ is created across the benzene molecule, which induces electron transmission
from the source lead to the drain lead, resulting in the tilted-band energy spectrum \cite{ref5},
shown in Figure \ref{fig6}(b).
An important feature is the localization length $L$, over which at least one occupied state's
wave function must remain delocalized across the entire benzene molecule.
Such a situation ensures that electron transport between the leads is then guaranteed, thus,
securing the integrity of the system.
In other words, to achieve this outcome, it is necessary for the energy of the highest occupied state in the drain lead
to always lie within the source-lead band, which is a condition requiring that its Fermi level $E_F$ stays
within that band, whence, by referring to Figure \ref{fig6}(b), we have that
\begin{equation}
E_F^{DL} \ge \alpha + 2 \beta,
\label{eq45}
\end{equation}
from which the maximum field gradient is seen to be
\begin{equation}
\tan \omega \equiv f_{max} = 2|\beta|/L,
\label{eq46}
\end{equation}
so that the valid range of $f$-values is given by
\begin{equation}
0 \le f \le 2|\beta|/L.
\label{eq47}
\end{equation}

\section{Results and Discussion}

Using the work of the previous sections, we are now in a position to calculate the transmission probability
function $T(E)$, by utilizing that form previously derived \cite{ref13} for transmission through a general
dimer impurity, namely,
\begin{equation}
T(E) = {{(1+2\gamma)^2 (4-X^2)} \over {(1-2Q)^2 (4-X^2) + 4(P-QX)^2}}  ,
\label{eq51}
\end{equation}
where
\begin{equation}
P = z_i+z_j  ~,~ Q = z_i z_j - \gamma -\gamma^2 ,
\label{eq52}
\end{equation}
with
\begin{equation}
z_{i,j} = (\tilde{\alpha}_{i,j} - \alpha)/ 2\beta ~,~ \gamma = (\tilde{\beta}_{ij} -\beta)/ 2\beta ,
\label{eq53}
\end{equation}
and $X$ given by (\ref{eq8}). Here, the renormalized parameters $\tilde{\alpha}_i$, $\tilde{\alpha}_j$
and $\tilde{\beta}_{ij}$ are chosen, from those presented in Sections 3 and 4, as appropriate for the type
of benzene dimer and presence of field. 
Parameter values are taken to be $\alpha = -6.5$ eV and $\beta = -2.7$ eV \cite{ref16}.

\begin{figure}[htbp]
\includegraphics[width=12cm]{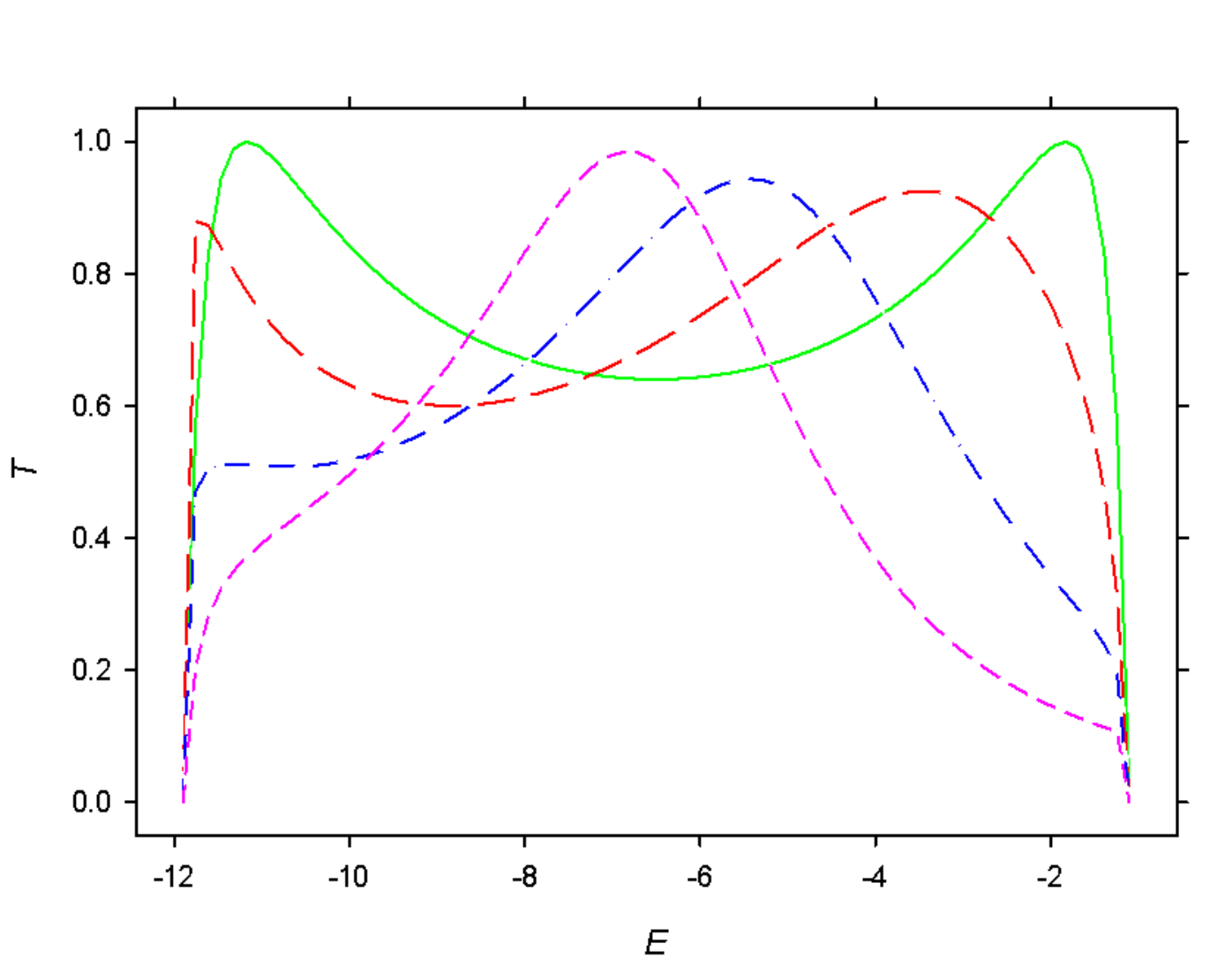}
\caption{Transmission $T$ versus energy $E$ for para-benzene, with field $f=$ (a) 0 (green solid curve), 
(b) 0.5 (red long-dashed), (c) 1.0 (blue dash-dotted), (d) 1.5 (pink short-dashed). }
\label{fig7}
\end{figure}

Looking first at para-benzene, consideration of  Figure \ref{fig3} indicates that $L=4a$, where $a=0.7 {\buildrel _{\circ} \over {\mathrm{A}}}$, from which condition (\ref{eq47}) leads to $0 \le f \le 1.93 ~\rm{eV}/{\buildrel _{\circ} \over {\mathrm{A}}}$ as 
the valid range of values for $f$. 
We start with the zero-field ($f=0$) case, shown in Figure \ref{fig7}(a) \cite{ref13}.
We observe that the $T(E)$ curve is symmetrical about the band center (at $E= \alpha = -6.5$), for which the curve has
a local minimum, then rising to a pair of local maxima, which are in fact resonances ($T=1$) at $E=\alpha \pm \sqrt 3 \beta
= -1.823$ and $-11.177$ eV. $T$ then drops to 0 at the band edges, $E= \alpha \pm 2 \beta = -1.1$ and $-11.9$ eV. 
There are no anti-resonances, for which $T=0$, within the band itself.
As soon as the field is switched on, so that $f>0$, the symmetry of the $T(E)$ curve is destroyed.
For a small field, such as is shown in Figure \ref{fig7}(b) for $f=0.5$, the $T(E)$ curve still bears close
resemblance to that for $f=0$, but with the two maxima reduced in height and shifted to lower energies,
the latter because the field gradient is $-f < 0$.
As $f$ increases further (see Figure \ref{fig7}(c) for $f=1.0$), the asymmetry becomes even more evident,
as the lower maximum is pushed towards the lower band edge, dropping in height as it does so.
Simultaneously, the upper maximum regains its height, becoming close to resonant, while still shifting to
lower energies.
These features are reinforced as $f$ increases even more (Figure \ref{fig7}(d) for $f=1.5$), with the lower
maximum disappearing completely and the upper maximum attaining a height of $T \approx 1$.
The shape of the $T(E)$ curve, shown in Figure \ref{fig7}(d), is very close to that for the limiting value of
$f=1.93$.
Overall, the general effect of the field is to enhance transmission at middling energies, while suppressing
it at higher and lower energies.

\begin{figure}[htbp]
\includegraphics[width=12cm]{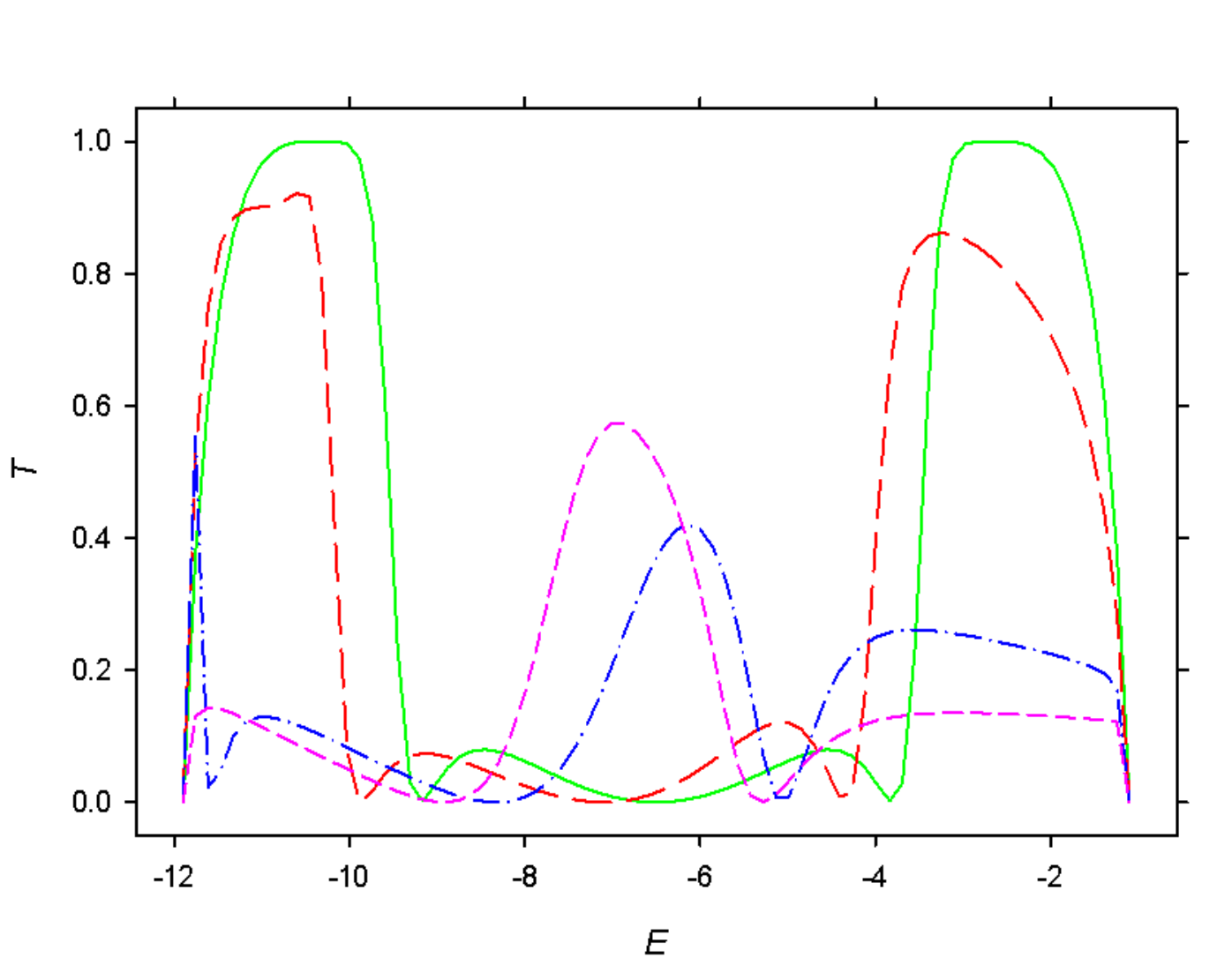}
\caption{Transmission $T$ versus energy $E$ for meta-benzene, with field $f=$ (a) 0 (green solid curve), 
(b) 0.5 (red long-dashed), (c) 1.5 (blue dash-dotted), (d) 2.0 (pink short-dashed). }
\label{fig8}
\end{figure}

Turning next to meta-benzene, Figure \ref{fig4} shows a value of $L=2 \sqrt 3 a$, which using (\ref{eq47})
produces $0 \le f \le 2.23 ~\rm{eV}/{\buildrel _{\circ} \over {\mathrm{A}}}$ as the physical range of $f$-values.
For the zero-field situation, shown in Figure \ref{fig8}(a)  \cite{ref13}, the $T(E)$ curve is symmetrical about the band
center (as was the case in Figure \ref{fig7}(a)), with anti-resonances occuring at $E=\alpha$ and $E=\alpha \pm \beta 
= -3.8$ and $-9.2$ eV.
The $T(E)$ curve also displays 4 maxima, 2 of them being resonances located at $E=\alpha \pm \sqrt 2 \beta = -2.7$
and $-10.3$ eV, and the other 2 being at $E=\alpha \pm (\sqrt 3 -2) \beta = -4.5$ and $-8.5$ eV.
As in the para- case, the symmetry occurs only for $f=0$. 
For a small field, such as $f=0.5$ in Figure \ref{fig8}(b), the basic shape of the $T(E)$ curve is similar, despite 
the breaking of  the symmetry. 
However, the anti-resonances and the maxima are all shifted to lower energies, with the resonances dropping in
height to $T < 1$. 
Of the two smaller maxima, the one at lower energy remains of roughly the same height, while the one at higher
energy increases its height modestly. 
These trends continue as $f$ is further increased (see Figure \ref{fig8}(c) for $f=1.5$), with again all anti-resonances
and maxima moving to lower energies. 
The two former resonances continue to shrink, while the height of the middle maxima continues to grow, and
becomes the dominant feature of the curve.
Moreover, the movement of the anti-resonances has the effect of narrowing the lowest sub-band, as its
bounding anti-resonance is pushed down towards the lower band edge.
The highest sub-band, which was initially dominant, is greatly diminished in height, but is somewhat broadened
due to the movement of its bounding anti-resonance.
With a further increase in $f$ to 2.0 (Figure \ref{fig8}(d)), these features are reinforced. 
In particular, the lowest original sub-band has now disappeared completely leaving a 3-sub-band structure, very
dissimilar to that seen for $f=0$, due to the broadening and changing heights of the remaining sub-bands.
Overall, the field effect is to increase transmission at middle energies, while decreasing it at the upper and lower
energies that originally housed resonances.

\begin{figure}[t]
\includegraphics[width=12cm]{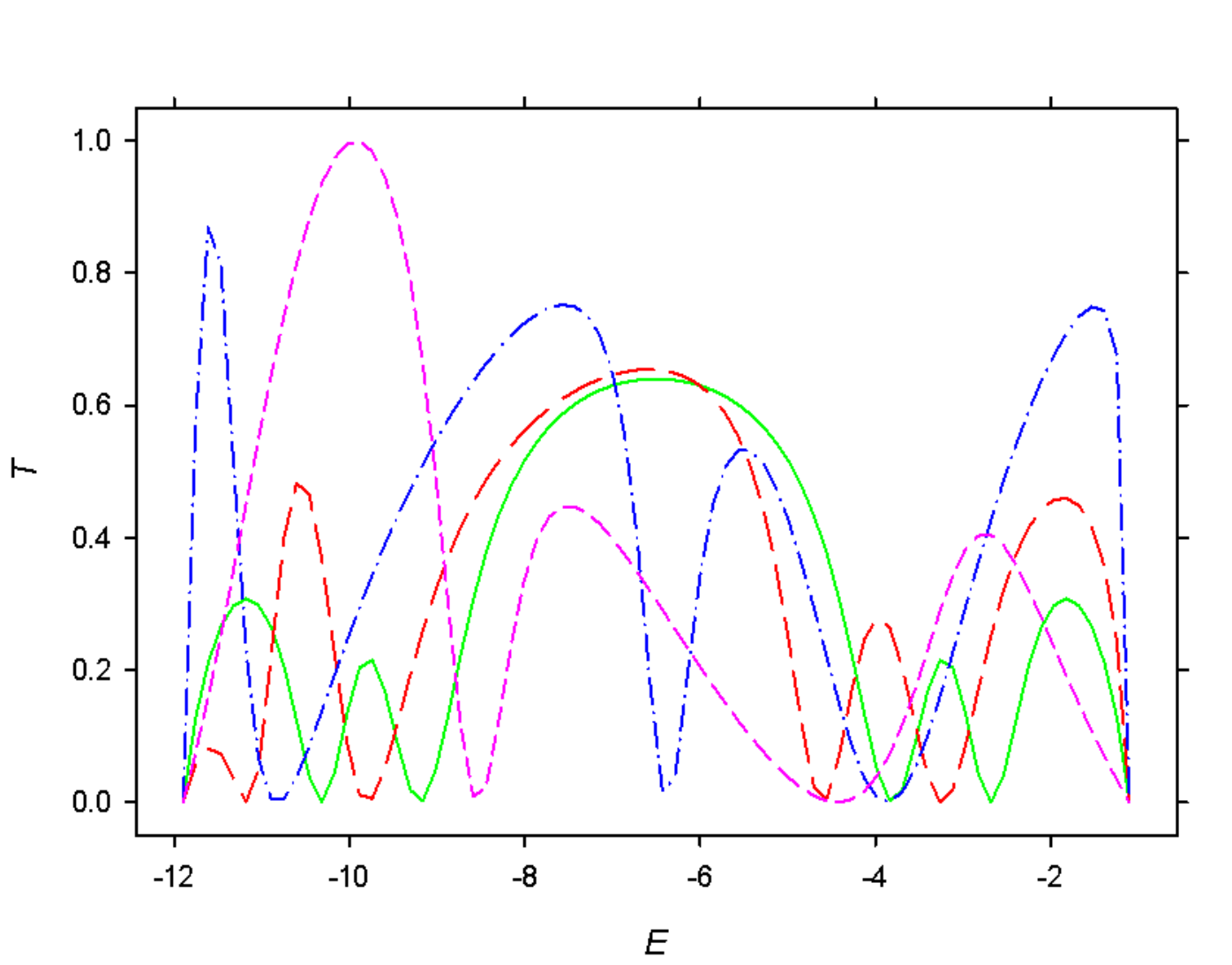}
\caption{Transmission $T$ versus energy $E$ for ortho-benzene, with field $f=$ (a) 0 (green solid curve), 
(b) 0.5 (red long-dashed), (c) 1.5 (blue dash-dotted), (d) 3.0 (pink short-dashed). }
\label{fig9}
\end{figure}

Lastly, we examine ortho-benzene, for which Figure \ref{fig5} gives $L=2a$, from which it follows, from (\ref{eq47}), that
$0 \le f \le 3.86 ~\rm{eV}/{\buildrel _{\circ} \over {\mathrm{A}}}$.
For the case of $f=0$ (Figure \ref{fig9}(a)), the $T(E)$ curve is once again symmetric about $E=\alpha$, with 
5 maxima (none of them resonances), which are positioned at $E = \alpha = -6.5$ eV, 
$E = \alpha \pm 2 \beta (0.609) = -3.21$ and $-9.79$
and $E=\alpha \pm \sqrt 3 \beta = -1.823$ and $-11.177$ \cite{ref17}.
There are 4 anti-resonances, located at $E=\alpha \pm \beta 
= -3.8$ and $-9.2$ eV and $E=\alpha \pm \sqrt 2 \beta = -2.7$ and $-10.3$ eV \cite{ref17}.
As in the two previous cases, switching on the field immediately breaks the symmetry of the curve, while
shifting anti-resonances and the maxima to lower energies.
For a small field (see Figure \ref{fig9}(b) with $f=0.5$), much of the effect is rather modest, although
most noticeably, the lowest sub-band is already showing substantial diminishment, both in height and width.
As the field is strengthened (see Figure \ref{fig9}(c) with $f=1.5$), this lowest sub-band continues to shrink
and eventually disappears, as its bounding anti-resonance is shifted so as to merge with the lower band edge.
The higher sub-bands actually show significant strengthening, in height and width.
Meanwhile, the lowest remaining sub-band is narrowed and poised to disappear, which it does
as $f$ increases further to $f=3.0$ (Figure \ref{fig9}(d)).
The sub-band that, for $f=0$, was the middle one, remains dominant for all $f$, and indeed becomes
temporarily resonant for $f \approx 3$.
The other two surviving sub-bands also show remarkable enhancement from their zero-field situation.
The net effect of the field is that transmission is marginally lowered at middle energies, while being
considerably elevated at higher and (especially) lower energies.

\section{Conclusions}

In summary, we have presented a model of electron transmission through a benzene molecule, subject
to an applied electric field. 
The $T(E)$ curves for all three benzene-leads configurations are symmetric in the 
zero-field case, but this symmetry is broken for any non-zero field strength, however small.
For each of the three scenarios, the variation in the $T(E)$ curve was examined, as the field gradient $f$ increases, 
with a general feature being a shift in both anti-resonances and peaks towards lower energies.

\section{Keywords}

molecular electronics, electron transmission, benzene, applied field, renormalization method


\begin{thebibliography}{99}

\bibitem{ref1} C. Zener, {\em Proc. Roy. Soc. (London)} {\bf 1934}, {\em 145}, 523.
\bibitem{ref2} H.M. James, {\em Phys. Rev.} {\bf 1949}, {\em 76}, 1611.
\bibitem{ref3} S. Katsura, T. Hatta, A. Morita, {\em Sci. Rep. T\^{o}huku Imp. Univ.} {\bf 1950}, {\em 35}, 19.
\bibitem{ref4} G. Wannier, {\em Phys. Rev.} {\bf 1960}, {\em 117}, 432.
\bibitem{ref5a} S.G. Davison, R.A. English, Z.L. Mi\v{s}kovi\'{c}, F.O. Goodman, A.T. Amos, B.L. Burrows, {\em J. Phys.:
Cond. Matt} {\bf 1997}, {\em 9}, 6371.
\bibitem{ref5}  S.G. Davison, K.W. Sulston, {\em Green-Function Theory of Chemisorption}, Springer, {\bf 2006}, ch. 7.
\bibitem{ref6} A. Aviram, M.A. Ratner, {\em Chem. Phys. Letts.} {\bf 1974}, {\em 29}, 277.
\bibitem{ref7} M.A. Reed, J.N. Randall, R.J. Aggarwal, R.J. Matyi, T.M. Moore, A.E. Wetsel, {\em Phys. Rev. Letts.} {\bf 1988},
{\em 60}, 535.
\bibitem{ref8} M. Di Ventra, S.T. Pantelides, N.D. Lang, {\em Appl. Phys. Letts.} {\bf 2000}, {\em 76}, 3448.
\bibitem{ref9} M.H. Hettler, W. Wenzel, M.R. Wegewijs, H. Schoeller, {\em Phys. Rev. Letts.} {\bf 2003}, {\em 90}, 076805.
\bibitem{ref10} Y.C. Choi, W.Y. Kim, K.-S. Park, P. Tarakeshwar, K.S. Kim, T.-S. Kim, J.Y. Lee, {\em J. Chem. Phys.}
{\bf 2005}, {\em 122}, 094706.
\bibitem{ref11} W.Y. Kim, K.S. Kim, {\em Acct. Chem. Res.} {\bf 2010}, {\em 43}, 111.
\bibitem{ref12}  R. Landauer, {\em IBM J. Res. Dev.} {\bf 1957}, {\em 1},  233;
{\em Phys. Lett. A.} {\bf 1981}, {\em 85},  91.
\bibitem{ref13} K.W. Sulston, S.G. Davison, {\em arXiv} {\bf 2015}, 1505.03808.
\bibitem{ref14} J.N. Murrell, S.F.A. Kettle, J.M. Tedder, {\em The Chemical Bond}, 2nd ed.,  Wiley, {\bf 1985}, ch. 9.
\bibitem{ref15} S.G. Davison, M. St\c{e}\'{s}licka, {\em Basic Theory of Surface States}, Oxford, {\bf 1992}, sect. 2.3.2.
\bibitem{ref16}  J.P. Lowe, K.A. Peterson, {\em Quantum Chemistry}, 3rd ed., Elsevier, {\bf 2006}, p. 278.
\bibitem{ref17}  K.W. Sulston, S.G. Davison, {\em arXiv} {\bf 2016}, 1607.05945 .

\end{thebibliography}
\end{document}